# Accelerating Atomistic Simulations with Piecewise Machine Learned Ab Initio Potentials at Classical Force Field-like Cost

Yaolong Zhang, Ce Hu, and Bin Jiang[*]

*Hefei National Laboratory for Physical Science at the Microscale, Key Laboratory of Surface and Interface Chemistry and Energy Catalysis of Anhui Higher Education Institutes, Department of Chemical Physics, University of Science and Technology of China, Hefei, Anhui 230026, China*

[*]: corresponding author: bjiangch@ustc.edu.cn

**Abstract**

Machine learning methods have nowadays become easy-to-use tools for constructing high-dimensional interatomic potentials with ab initio accuracy. Although machine learned interatomic potentials are generally orders of magnitude faster than first-principles calculations, they remain much slower than classical force fields, at the price of using more complex structural descriptors. To bridge this efficiency gap, we propose an embedded atom neural network approach with simple piecewise switching function based descriptors, resulting in a favorable linear scaling with the number of neighbor atoms. Numerical examples validate that this piecewise machine learning model can be over an order of magnitude faster than various popular machine learned potentials with comparable accuracy for both metallic and covalent materials, approaching the speed of the fastest embedded atom method (*i.e.* several $\mu$s/atom per CPU core). The extreme efficiency of this approach promises its potential in first-principles atomistic simulations of very large systems and/or in long timescale.

# Introduction

Atomic-scale molecular dynamics (MD) and Monte Carlo (MC) simulations are powerful computational tools for studying dynamic and thermodynamic properties of molecules, chemical reactions, and materials. Potential energy surface (PES) is a fundamental ingredient for atomistic simulations. A desirable PES should be not only sufficiently accurate to reproduce ab initio energies and atomic forces, but also highly efficient to enable large-scale repeated calculations. For macromolecules and materials, where ab initio calculations are generally infeasible, physically motivated empirical or semi-empirical force fields[1-5], have extensively been used over the years. Despite their limited accuracy, these classical force fields (CFFs) are extremely fast due to their simple analytical forms, which afford MD/MC simulations with millions of atoms and nanosecond timescale[6].

Recent years have witnessed the fast popularization of machine learning (ML) in many fields[7, 8]. In particular, ML approaches have emerged in constructing PESs of systems across gas[9-15] and condensed phase[16-21], commonly referred as machine learned interatomic potentials (MLIPs). In particular, to deal with high-dimensional systems, most MLIPs express the total energy of the system as the sum of atomic energies like CFFs[22]. Instead of relying on physically-derived functions, MLIPs are able to learn the relationship between the atomic local environment and the atomic energy given a set of ab initio data points[22]. Thanks to the more complex and flexible mathematic forms, these MLIPs are intrinsically more accurate than CFFs and able to reproduce ab initio molecular dynamics (AIMD) results at a fraction of cost[23]. We note that it remains challenging to construct a PES for complex biomolecular systems, due partly to its difficulty in obtaining accurate ab initio energies and forces.

Despite these successes, an important fact attracting less attention is that these MLIPs are still much more expensive than CFFs, at the price of the increasing number of parameters. As pointed out in a recent perspective[23], MLIPs for typical periodic systems have been reported to run per MD step (*i.e.* compute atomic forces once) at a speed of about $10^{-4} \sim 10^{-2}$ s/atom on a single CPU core[6, 19, 20, 24, 25], which are at least two orders of magnitude slower than the fastest embedded atom method (EAM) force field[6, 24]. The computational bottleneck of most MLIPs is the evaluation of structural descriptors that distinguish local atomic configurations[26]. These descriptors in general need to sum over many-body interactions between the central atom and its neighbors so as to preserve the rotational, translational, and permutational symmetry of the system[22]. As a result, the numerical cost of conventional descriptors, for example, the Behler-Parrinello type atom-centered symmetry functions (ACSFs)[16] and their variants[26-28], often scales at least quadratically with the number of neighbor atoms. Alternative methods have been proposed by several groups based on direct expansion of the atomic energy in various forms in terms of symmetry-invariant many-body basis functions[29] (*e.g.*, polynomials[19, 30] or bispectrum components[18]). These methods rely on standard (mostly linear) least squares optimization of expansion coefficients, invoking neither neural networks (NNs) nor kernel-based regression. Linear scaling with respect to the number of neighbor atoms is achieved by converting the sum over many-body terms into a product of two-body sums[18, 19, 29, 30].

From a different perspective, we have recently derived a linear-scaling ML framework inspired by the EAM concept. This so-called embedded atom neural network (EANN) model[31] combines the implicit description of three-body interactions by two-body terms with the flexibility of NNs. In this work, we replace the original Gaussian-type orbitals with piecewise switching functions in the construction of structural descriptors. This greatly lowers the scaling factor and improves the efficiency of the EANN model. Tests on representative periodic systems demonstrate that the piecewise EANN (PEANN) potentials can be over one order of magnitude faster than various well-established MLIPs with comparable accuracy, and in some cases as fast as several μs/atom/CPU-core that is reachable by simplest CFFs only before. This PEANN model will offer a promising solution to the dilemma of accuracy versus efficiency of MLIPs in very large systems.

**Method**

We shall first review the original EANN model briefly, which borrows the idea of EAM expressing the atomic energy as a functional of the embedded electron density of the impurity atom, *i.e.* $E_i = \mathcal{F}[\rho(\hat{\mathbf{r}}_i)]$ [1]. In particular, we replace the scalar density value $\rho(\hat{\mathbf{r}}_i)$ with a set of density-like descriptors made of atomic orbitals in the vicinity of embedded atom and the complicated functional $\mathcal{F}$ with an atomic NN[31]. In this regard, any type of atomic orbitals can be taken as long as they effectively distinguish the local environment. For simplicity, in the original EANN approach[31-34], we use the Gaussian-type orbital (GTO) in Cartesian space in the following form,

$$\varphi_{l_x l_y l_z}^{\alpha, r_s}(\hat{\mathbf{r}}) = x^{l_x} y^{l_y} z^{l_z} \exp\left[-\alpha(r - r_s)^2\right], \tag{1}$$

where $\hat{\mathbf{r}} = (x, y, z)$ represents the position vector of an electron relative to the nucleus, $r = |\hat{\mathbf{r}}|$, $\alpha$ and $r_s$ are hyperparameters that determine widths and centers of Gaussian radial functions, $l_x + l_y + l_z = L$ specifies the orbital angular momentum ($L$). To find the electron density of the embedded atom $i$ at location $\hat{\mathbf{r}}_i$, we evaluate individual electron density contributions by the square of linear combination of these atomic orbitals centered at surrounding atoms with the same $L$, $\alpha$ and $r_s$,

$$\rho_{L,\alpha,r_s}^i = \sum_{l_x,l_y,l_z}^{l_x+l_y+l_z=L} \frac{L!}{l_x! l_y! l_z!} \left[\sum_{j=1} c_j \varphi_{l_x l_y l_z}^{\alpha, r_s}(\hat{\mathbf{r}}_i - \hat{\mathbf{r}}_j) f_c(r_{ij})\right]^2, \tag{2}$$

In Eq. (2), the summation in spirit mimics the integral over the embedded wavefunction in terms of atomic orbitals of nearby atoms within a cutoff radius ($r_c$), with $c_j$ being the corresponding adjustable expansion coefficient of atom $j$ at location $\hat{\mathbf{r}}_j$ which is optimized in the training process and $f_c(r_{ij})$ a cutoff function[35] to ensure that the contribution of each neighbor atom decays smoothly to zero at $r_c$,

$$f_c(r_{ij}) = \begin{cases} [0.5 + 0.5\cos(\pi r_{ij}/r_c)]^2, & r_{ij} \leq r_c \\ 0, & r_{ij} > r_c \end{cases}, \tag{3}$$

where $r_{ij} = |\hat{\mathbf{r}}_i - \hat{\mathbf{r}}_j|$ is the internuclear distance between atom $i$ and atom $j$. A key advantage of these density-like descriptors is that they can be formally transformed to a series of angular basis[30], preserving the invariance of translation, rotation and permutation. In this way, they unify the radial and angular functions as defined in conventional ACSFs with an implicit description of three-body interactions, realizing a linear scaling with respect to the number of neighbors.

The remaining cost of computing these descriptors results dominantly from the evaluation of the Gaussian radial function and the cutoff function. In particular, one has to explicitly calculate Gaussian radial function and $f_c(r_{ij})$ for any atom $j$ inside the cutoff sphere for a given set of hyperparameters. This becomes actually wasteful when Gaussian radial function and $f_c(r_{ij})$ almost vanish (*e.g.*, $r_{ij}$ deviates significantly from $r_s$), thus drastically more expensive with the increasing number of descriptors and cutoff radius. To overcome this shortcoming, we replace the Gaussian radial function with a simple piecewise switching function,

$$f^{\alpha}_{r_{in},r_{out}}(r) = \begin{cases} 1, & \beta \leq 0 \\ \dfrac{exp\left[-\alpha x^2(2-\beta)^2\right]-exp(-\alpha)}{1-exp(-\alpha)}, & 0 < \beta < 1, \\ 0, & \beta \geq 1 \end{cases} \quad (4)$$

which is characterized by its inner ($r_{in}$) and outer ($r_{out}$) ends with $\beta = (r - r_{in})/(r_{out} - r_{in})$ and the damping strength ($\alpha$). This piecewise switching function is carefully designed to warrant its continuousness up to the first order derivative with respect to $r$ at the boundary of $r_{in}$ and $r_{out}$. The piecewise atomic orbital (PAO) then becomes,

$$\varphi^{\alpha,r_{in},r_{out}}_{l_x,l_y,l_z}(\hat{\mathbf{r}}) = x^{l_x} y^{l_y} z^{l_z} f^{\alpha}_{r_{in},r_{out}}(r), \quad (5)$$

This replacement has several distinct advantages. First, $f^{\alpha}_{r_{in},r_{out}}(r_{ij})$ is now computed only when the $j$th atom is at the distance of $r_{in} < r_{ij} < r_{out}$ to the central atom $i$, otherwise its value is simply zero or one. Second, defining $r_{out} = r_c$ in the last piecewise function, $f^{\alpha}_{r_{in},r_{out}}(r_{ij})$ naturally goes to zero at $r_c$. Consequently, the artificial cutoff function $f_c(r_{ij})$ is no longer necessary. These features are best illustrated in Fig. 1, which compares the radial distributions of a group of normalized GTOs and PAOs, where the angular part has been factorized out leaving a factor $r^L$ (see Supplementary

Information for details). As seen in Fig. 1a, Gaussian radial functions are expanded in the full range within $r_c$ with their centers shifted incrementally. They need to be explicitly calculated whatever $r_{ij}$ is even if their contributions are negligible. Similar procedure is indeed unavoidable in most MLIPs. In comparison, as displayed in Fig. 1b, each piecewise radial function is a continuous function with its the left half domain given by the factor $r^L$ and its right half domain switching from one to zero. Here $r^L$ is an auxiliary factor for illustration and transformation to explicit angular basis only, but virtually not computed. As a consequence, each $f^{\alpha}_{r_{in},r_{out}}(r_{ij})$ function needs to be calculated once provided that $r_{ij}$ falls in one of these intervals, namely $[r_{in}, r_{out}]$, giving rise to significant savings. We note that a similar choice of piecewise cosine functions has been proposed by Huang et al.[36] in their single atom neural network model. However, those cosine functions behave similarly as the GTOs, which need to be explicitly computed within twice the interval of $f^{\alpha}_{r_{in},r_{out}}(r_{ij})$ here. Singraber et al. also proposed to use an exponential function as a cutoff function in their NN implementation[37]. In this work, we have learned both energies and forces simultaneously in the training process. The forces of PEANN potentials are obtained analytically by applying the chain rules of derivatives from NN output to descriptors, and then to Cartesian coordinates.

## Results

In spite of a large body of ML approaches for constructing PESs, very limited studies have systematically assessed their relative efficiency when reaching a similar accuracy based on the same benchmark data set and the optimal number of parameters[24, 38-40]. This actually prevents a direct comparison of efficiency among various MLIPs. In the present work, we first take these freely available data reported by Ong and coworkers for Cu, Mo, and Ge systems[24]. These systems span various crystal structures (fcc, bcc, and diamond) and bonding types (metallic and covalent), serving as great tests for the universality of a ML model. For each element, a few hundreds of structures were sampled via high temperature AIMD simulations of different bulk supercells at density functional theory (DFT) level, along with strained structures with varying cell sizes and high-index surface structures. These very diverse structures cover a huge configuration space with distinct atomic local environments, providing stringent challenges for ML models. More importantly, the performance and computational cost of several popular MLIPs and CFFs have been benchmarked in Ref. 24 given the same datasets, offering us useful references. We also used exactly the same training and test sets for these systems in learning our EANN and PEANN models. Periodic boundary conditions were fulfilled in the PESs. Technical details on training the EANN and PEANN models are given in the Electronic Supplementary Information (ESI).

In Table 1, we compare the root-mean-square-errors (RMSEs) in energy and atomic force in the test set of various MLIPs and CFFs (whenever available), as well as the corresponding computational costs per MD step with a single CPU core. Apparently, all MLIPs are more accurate than CFFs, especially with much lower RMSEs in energy. As discussed in Ref. [24], the Gaussian approximation potential (GAP)[17] and moment tensor potential (MTP)[19] have somewhat lower RMSEs, followed by several the NN-based potentials, and then the spectral neighbor analysis potential (SNAP)[18] and its quadratic version (qSNAP)[41]. Given the fact that those limited data points are diversely distributed, it is understandable that the kernel-based GAP and the invariant polynomial based MTP methods (like an atom centered local version of the permutationally invariant polynomial method[13]) are advantageous, which typically require fewer points than NN-based methods to converge the PES to reasonable level of accuracy[15]. NN-based and SNAP models can improve their accuracy with increasing number of training data[24] (*e.g.*, see also Ref. [31] for much better performance of EANN for bulk systems with more data available). Indeed, all MLIPs show excellent performance in predicting properties energies such as phonon dispersion curves (Fig. S1). To test the extrapolability of the EANN/PEANN potentials, taking Mo as an example, we have performed 250 ps NVT classical MD simulations for a 3×3×3 bulk Mo with a 0.1 fs time step at 1300K, as done in Ref. 24. (see ESI for details). In Fig S2, we further compare the prediction errors of these MLIPs for 40 snapshots extracted from these trajectories with a time interval of 2.5 ps, which have covered over a completely different chemical space from the training data. Overall, their accuracy are comparable and more or less converged with the number of basis functions/polynomials/kernels, allowing a fair comparison of their cost[24].

We shall note that the evaluation time of the same potential is dependent on the processor used. These numbers reported in this work were based on the implementation with a 28-core processor, Intel(R) Xeon 6132 2.60GHz, while those taken from other work have been scaled by an estimated factor for comparison whenever necessary. As seen in Table 1, it is clear that these GAPs run most slowly on the order of several milliseconds/atom/core, due apparently to the local interpolation nature of this kernel-based method. Other ML methods based on global fitting, including MTP, SNAP/qSNAP, and Behler-Parrinello NN (BPNN) models, are typically dozens of times faster than GAPs, at the speed of a few $10^{-5}$~$10^{-4}$ s/atom/core. Compared to these methods, the original EANN model speeds up by nearly one order of magnitude, reaching 6.5~12.0 μs/atom/core for the three systems. The PEANN model further lowers the cost to an unprecedented level, *e.g.*, ~4.9 μs/atom/core for bulk Cu, which is merely about twice that of the simplest EAM force field[42], yet one twelfths of that of the modified EAM potential[43]. This is quite encouraging as these EAM models[42-44] implemented in LAMMPS are highly optimized with the extensive use of tabulated values[6], while our PEANN potentials are so far implemented with an in-house Fortran MD code[45]. For Ge, where a Tersoff potential[46] has to be used instead of EAM due to its covalent bonding nature, the PEANN model is again only ~3 times more expensive (7.6 μs/atom/core). It is important to emphasize that our PEANN model, like other MLIPs, is equally suitable for both metallic and covalent systems and more accurate than corresponding CFFs. It should be noted that the current implementation of PEANN/EANN is based on the Linked-List cell algorithm[47] which may be further improved by a dynamic Verlet neighbor list[48] when interfaced with the sophisticated MD codes. It is worth noting that Mueller and coworkers recently constructed a new type of MLIP for copper based on symbolic regression that simultaneously optimizes some simple functional forms and their parameters using genetic algorithm[49]. Given its very simple expression, it achieved a comparable speed as the EAM potential for copper. Whereas it remains unclear about its universality to describe more directional bonding in molecules and/or reactive systems as other MLIPs.

The improved efficiency of PEANN over EANN arises solely from the decreasing cost of evaluating structural descriptors. In Table S1, we count the individual time for computing all internuclear distances (namely the first necessary step in any MLIP or CFF), constructing density-like descriptors based on internuclear distances (the second step), as well as evaluating neural networks (the third step), respectively. Taking Cu as an example, the consuming time for descriptors is reduced from 3.5 in EANN to 1.9 μs/atom/core in PEANN. Put another way, replacing the same number of GTOs with PAOs virtually speeds up by roughly a factor of two, because almost half of PAOs are not explicitly computed as seen in Fig. 1. This actually represents a significant improvement on the basis of the sufficiently efficient EANN model. Nevertheless, such an improvement does not significantly reduce the total cost of the PEANN potential of Cu, as computing internuclear distances and NNs is now as expensive as computing structural descriptors. However, once we artificially raise the number of descriptors or the cutoff radius (equivalently the average number of neighbor atoms), as illustrated in Fig. 2, the computational costs of the PEANN and EANN potentials for Cu both increase linearly, while the former becomes increasingly more efficient than latter. These results not only prove the linear scaling behavior of both models, but also suggest the superiority of the PEANN model in more complex systems that require a larger cutoff radius and a larger number of descriptors.

To validate this point and show the generalizability of the PEANN model, we construct EANN and PEANN potentials for bulk water, which is an important condensed phase system that has been extensively studied with many ML methods[20, 38, 50-55]. Again, for our purpose, we choose to fit the ab initio data reported by Cheng *et al.*[54], consisting of energies and forces for 1,593 diverse structures of 64 molecules of liquid water at the revPBE0-D3 level[56, 57]. These data have been used in that work to construct a public BPNN potential[16], allowing us to make direct comparison. The BPNN potential gives the test RMSEs of 2.3 meV/atom (energy) and 120 meV/Å (force)[54], which are comparable to the values of 2.1 meV/atom (energy) and 129meV/Å (force) given by the EANN model, and 2.3 meV/atom (energy) and 131 meV/Å (force) by the PEANN model. Fig. 3 compares the O-O, O-H, and H-H radial distribution functions (RDFs) obtained via 20 ps NVT classical MD simulations using these NN potentials (detailed in ESI). The excellent agreement suggests that these models achieve comparable level of accuracy so that their computational costs can be reasonably compared in Fig. 4 as a function of the number of atoms. As expected, all MLIPs scale linearly with respect to the number of atoms, which is more favorable than the cubic scaling of DFT calculations. Specifically, the cost of the BPNN potential is $\sim 4.1 \times 10^{-4}$ s/atom/core, which is roughly five times that of the EANN model ($\sim 7.8 \times 10^{-5}$ s/atom/core) and 12 times that of the PEANN model ($\sim 3.4 \times 10^{-5}$ s/atom/core). For reference, we find that available data about the costs of other NN potentials[20, 37, 58] of water trained with different data sets but with a similar cutoff radius (*i.e.* $r_c \approx 6$ Å, which largely determines the efficiency of a MLIP) are on the order of $10^{-4} \sim 10^{-3}$ s/atom/core, no better than the BPNN potential of Cheng *et al.*[54]

Also compared in Figs. 3 and 4 is the well-known TIP4P force field[59] which is based on physically motivated functions to represent the interactions of rigid water molecules. Given the simplistic form of the TIP4P force field, it is not surprising that this empirical model runs more efficiently than all ab initio MLIPs, reaching $\sim 1.0\times 10^{-5}$ s/atom/core. We note that more accurate classical force fields exist, *e.g.* AMOEBA[5], which would be yet more complicated and expensive than the TIP4P model. The efficiency of our PEANN model is expected to be closer to those more advanced force fields. Indeed, the more anisotropic covalent and hydrogen bonding in liquid water demands more complex descriptors and longer cutoff radii than those used in metallic systems, resulting in higher computational costs of MLIPs in general. However, these MLIPs are fully dissociable with faithful representation of diverse DFT data including vdW corrections yielding more accurate RDFs (see Fig. 3), among which the PEANN model is outstanding in the trade-off between efficiency and accuracy. Although this piecewise strategy allows us to extend the local environment to some extent, the long-range Coulomb interactions are still difficult to be described effectively, which depend on monomer properties such as dipole moment and polarizability[55] and deserve further exploration.

## Discussion and conclusion

To summarize, in this work, we demonstrate the high accuracy and scalability, universality, and most importantly the extremely low computational cost of the newly proposed PEANN model. The superior efficiency of the PEANN model results from the linear scaling with the number of neighbor atoms and the piecewise switching function based descriptors that are not necessarily computed for each neighbor atom. As illustrated in several benchmark systems, the PEANN model greatly closes the efficiency gap between typical MLIPs and physically motivated CFFs by at least an order of magnitude for both metallic and covalent systems. Impressively, the PEANN potentials for single component systems reach the speed of several μs/atom/core per step, as the EAM-type CFFs, meanwhile retain the ab initio accuracy as conventional MLIPs. This piecewise strategy can be also integrated with other existing ML models, which deserves further exploration. It should be emphasized that it remains challenging to develop MLIPs for complex biomolecular systems even with the proposed PEANN method, for which accurately computing ab initio energies and forces are extremely demanding. Further acceleration of the PEANN model may be achieved by linking with GPU-based deep neural network packages and an optimal implementation with sophisticated MD codes like LAMMPS[47], which would enable classical and/or path integral MD/MC simulations in complex systems to long timescale and to explore rare events. Work along these directions are in progress in our group.

**Conflict of interest:** The authors declare that they have no conflict of interest.

**Acknowledgements:** This work was supported by National Key R&D Program of China (2017YFA0303500), National Natural Science Foundation of China (22033007, 91645202, 21722306), and Anhui Initiative in Quantum Information Technologies (AHY090200). We appreciate the Supercomputing Center of USTC and AM-HPC for high-performance computing services. We thank Prof. Shyue Ping Ong and Yunxing Zuo for sending us extra data to help test our code.

**Data availability:** The EANN and PEANN potentials reported in this work are freely available at https://github.com/zylustc/Piecewise-embedded-atom-neural-networks-potential Data points used in this work for Cu, Ge, Mo, and water systems should be found from original publications, namely Ref. [24] and Ref. [54].

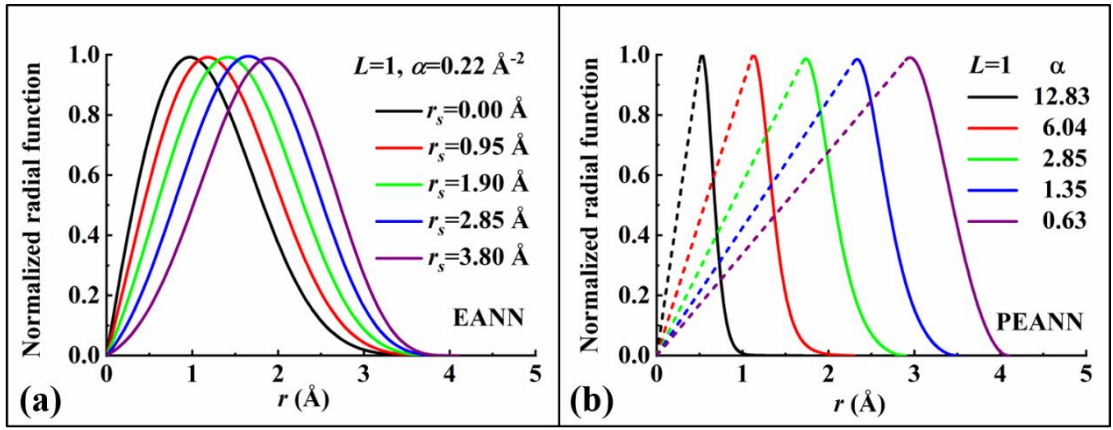

Figure 1. Radial functions extracted from structural descriptors of EANN and PEANN models of Cu. (a) EANN with five evenly distributed $r_s$ between 0~3.8 Å and (b) PEANN with five evenly distributed $r_{in}$ and $r_{out}$ grids within 0.5~2.9 Å and 1.7~4.1 Å, respectively.

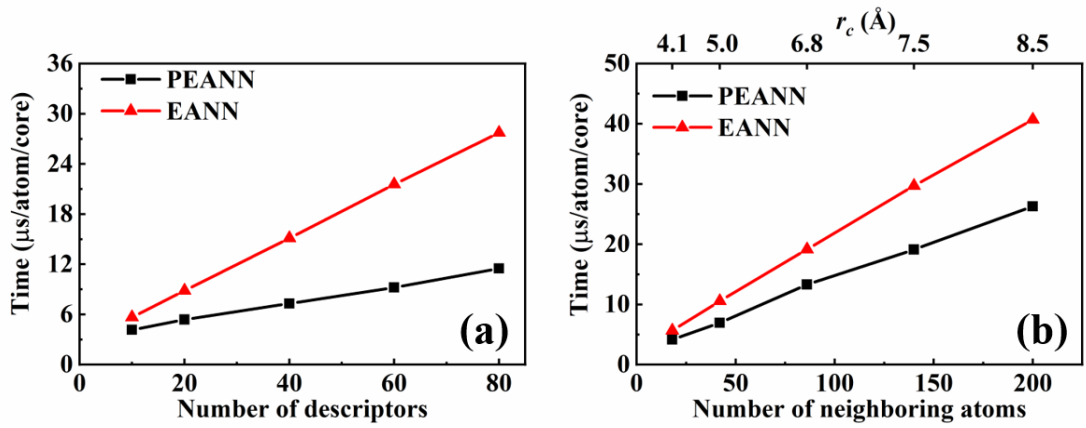

Figure 2. Computational cost of EANN and PEANN with respect to (a) the number of descriptors and (b) the number of neighboring atoms (corresponding $r_c$ shown on the top), respectively.

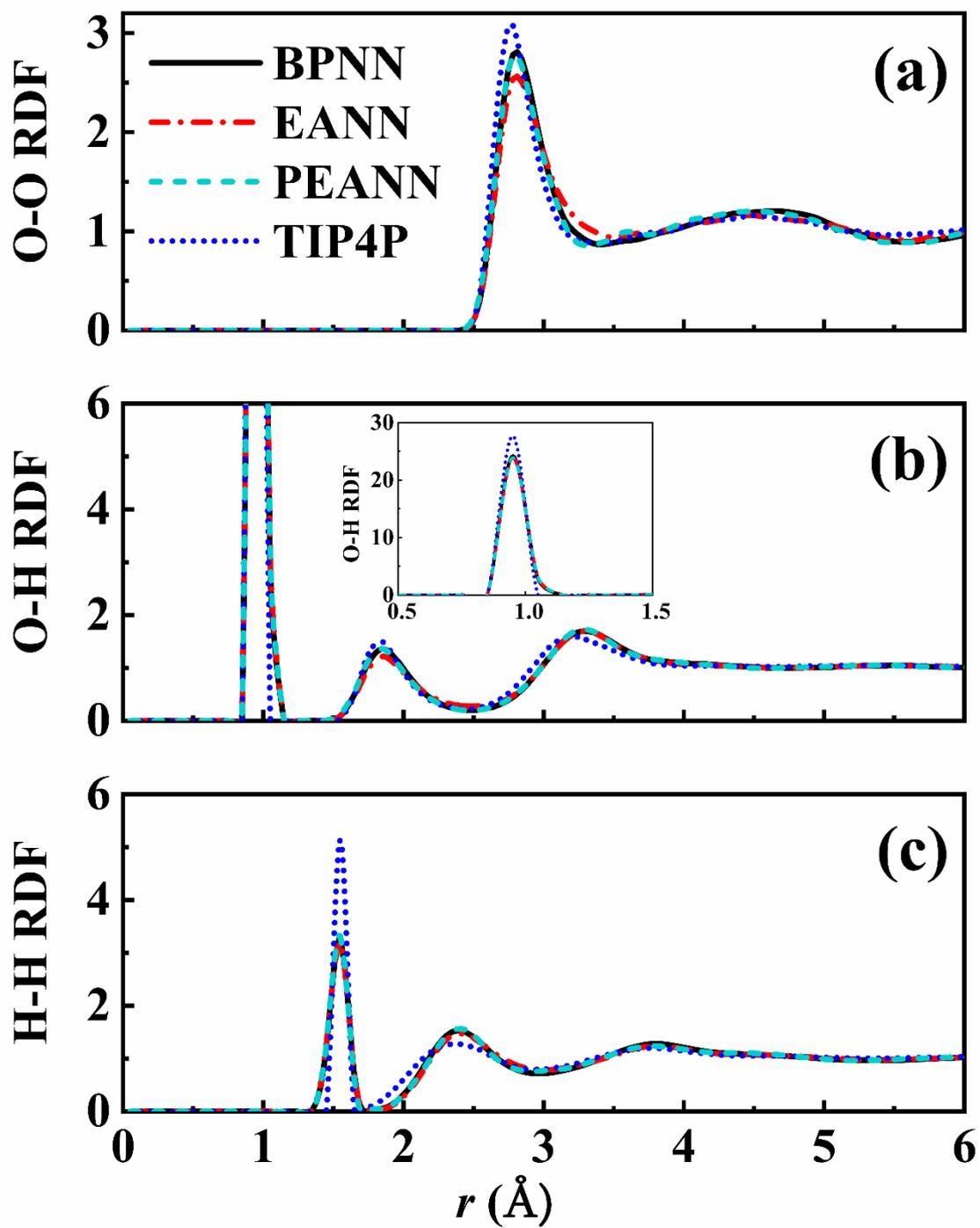

Figure 3. (a) O-O, (b) O-H and (c) H-H RDFs of liquid water obtained by MD simulation using BPNN, EANN, PEANN, and TIP4P potentials at 300K.

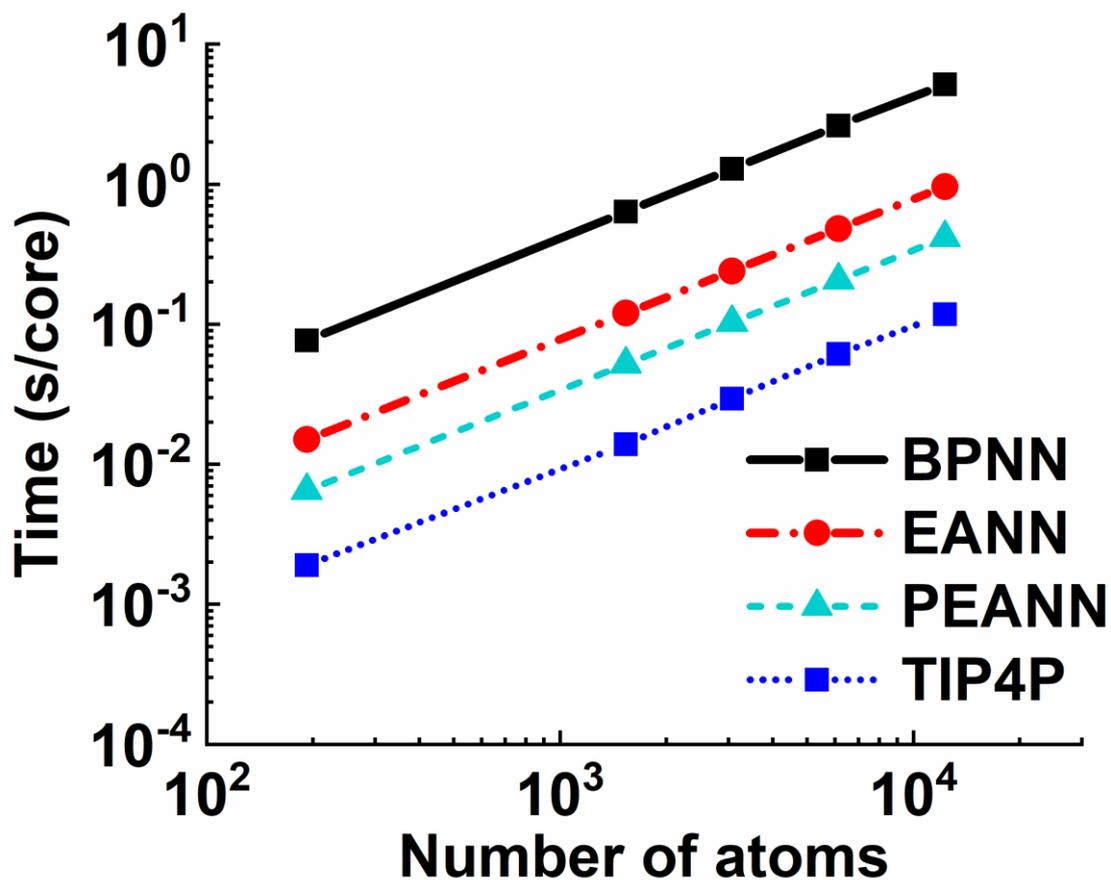

Figure 4. Computational cost per MD step with respect to the number of atoms in bulk water systems calculated with BPNN, EANN, PEANN and TIP4P models, based on the Intel(R) Xeon 6132 2.60GHz processor.

Table 1: Comparison of test RMSEs and computational costs per atom on a single processing core (μs/atom/core) of several MLIPs and CFFs for representative systems. RMSEs in energies and atomic forces (in parentheses) are in meV and meV/Å.

| System | Cu | | Ge | | Mo | |
|---|---|---|---|---|---|---|
| | RMSE | Time | RMSE | Time | RMSE | Time |
| PEANN | 0.7 (32) | 4.9 | 3.3 (105) | 7.6 | 4.2 (246) | 8.7 |
| EANN | 0.6 (26) | 6.5 | 3.0 (97) | 11.6 | 4.5 (245) | 12.0 |
| GAP | 0.6 (20) | 5349.4* | 4.5 (80) | 5555.6* | 3.6 (160) | 2654.7* |
| MTP | 0.5 (10) | 120.0* | 3.7 (70) | 369.5* | 3.9 (150) | 126.7* |
| BPNN | 1.7 (60) | 80.8* | 11.0 (120) | 109.7* | 5.7 (200) | 95.5* |
| SNAP | 0.9 (80) | 269.1* | 11.0 (290) | 332.5* | 9.1 (370) | 115.7* |
| QSNAP | 1.2 (50) | 94.2* | 10.6 (200) | 121.2* | 4.0 (330) | 115.7* |
| EAM | 7.5 (120) | 2.2 | \ | \ | 68.0 (520) | 1.6 |
| Tersoff | \ | \ | 550.4 (1360) | 2.4 | \ | \ |
| MEAM | 10.5 (240) | 61.3 | \ | | 36.4 (220) | 46.2 |

*These values were taken from Ref. [24] multiplied by a factor ~0.8 to account for the different intrinsic efficiency of Intel(R) Xeon 6132 2.60GHz (this work) versus Intel i7-6850k 3.6 GHz (in Ref. [24]) processors. The scaling factor is roughly estimated by the average time ratio of computing EAM, MEAM, and SANP potentials using both processors and its value would not alter the main conclusion here.

**Electronic Supplementary Information**

**Accelerating Atomistic Simulations with Piecewise Machine Learned Ab Initio Potentials at Classical Force Field-like Cost**

Yaolong Zhang, Ce Hu, and Bin Jiang*

*Hefei National Laboratory for Physical Science at the Microscale, Key Laboratory of Surface and Interface Chemistry and Energy Catalysis of Anhui Higher Education Institutes, Department of Chemical Physics, University of Science and Technology of China, Hefei, Anhui 230026, China*

*: corresponding author: bjiangch@ustc.edu.cn

# I. Details on training neural networks

In the embedded atom neural network (EANN) approach[1] and its piecewise version (PEANN), the weights and biases of atomic NNs along with the atomic expansion coefficients were determined by minimizing the cost function defined by root-mean-errors between ab initio potential energies and atomic forces with respect to Cartesian coordinates and corresponding NN outputs[2],

$$S(\mathbf{w}) = \sum_{i=1}^{N_{data}} [(E_i^{NN} - E_i^{Ref})^2 + \eta |\vec{\mathbf{F}}_i^{NN} - \vec{\mathbf{F}}_i^{Ref}|^2]/N_{data}. \qquad (S1)$$

Here, $\mathbf{w}$ is a collection of all adjustable parameters, $N_{data}$ is the number of configurations in the training set. $E_i^{NN}$, $E_i^{Ref}$, $\vec{\mathbf{F}}_i^{NN}$ and $\vec{\mathbf{F}}_i^{Ref}$ are potential energies and atomic force vectors of $i$th configuration obtained by NN and reference ab initio calculations, respectively. Note that each $E_i^{NN}$ and $\vec{\mathbf{F}}_i^{NN}$ are the sum of atomic NN outputs and for the atomic NN parameters are identical for same element. An efficient hybrid extreme machine learning Levenberg-Marquardt (ELM-LM) algorithm was employed to optimize these adjustable parameters[3].

For all systems discussed in this work, the NN structures consist of two hidden layers. To make a fair comparison, the number of neurons in each hidden layer, the number of descriptors, as well as the cutoff radius ($r_c$) were all kept the same for in EANN and PEANN potentials of each system. Table S1 gives such information for the four benchmark condensed phase systems, namely Cu, Ge, Mo, and water. Also compared in Table S1 are the computational costs of evaluating individually the inter-nuclear distances within $r_c$, the density-like structural descriptors, and NNs (plus the rest minor contributions), respectively. Other hyperparameters to determine the density-like descriptors are listed in Tables S1-S5 for PEANN and Tables S6-S9 for EANN, for Cu, Ge, Mo, and water in sequence. Note that for the water system, we have applied the CUR matrix decomposition algorithm[4] to select the optimal descriptors that best represent the training set, as used by Ceriotti and coworkers to optimize the selection of Behler-Parrinello type atom centered symmetry functions[5].

## II. Molecular dynamics simulations

To demonstrate the comparable performance of EANN and PEANN as other machine learning models, as done in Ref. 12, we have predicted material properties such as phonon dispersion curves, as well as energies and forces at unseen structures, taking Mo as an example (this is already the system with largest errors). To this end, in Fig. S1, we compare the DFT calculated phonon spectrum for 3×3×3 bulk Mo with the results predicted by all machine learning models discussed in the main text. In addition, following Zuo et al.[12], we have performed 250 ps NVT classical molecular dynamics (MD) simulations for a 3×3×3 bulk Mo with a 0.1 fs time step at 1300K maintained with Andersen thermostat, and extracted 40 snapshots with a time interval of 2.5 ps. It should be noted that none of the training and testing data were obtained from such high temperature MD simulations and the sampled configurations could be therefore very different and unknown, serving as good candidates for testing the extrapolability of machine learning potentials. The energy and force error distributions of our EANN and PEANN model are compared in Fig S2 with other machine learning models whose results are extracted from Ref. 12.

To validate the accuracy of our PEANN and EANN potentials for liquid water, we compare the O-O, O-H and H-H radial distribution functions (RDFs) of liquid water obtained by the classical molecular dynamics (MD) simulations using the TIP4P model in Ref. [6], Behler-Parrinello NN (BPNN) potential in Ref. 7, EANN and PEANN potentials in this work, respectively. MD simulations have been performed with 64 water molecules in a cubic box with its side length of 12.42 Å at temperature of 300 K. A total of 20 ps NVT MD simulations with a time step of 0.2 fs. The Andersen thermostat[8] was used for keeping the temperature in the simulations of the PEANN/EANN potentials in a modified VENUS code[9], while the Nose-Hoover Chains algorithm[10] was used for the TIP4P and BPNN models implemented with LAMMPS[11], respectively. It is found that the TIP4P force field requires a longer cutoff radius and long-range corrections to yield the correct description of RDFs as shown in Fig. 4, and our setup thus follows the original publication[6] where the short-range interactions were truncated at 7.75 Å.

Table S1: NN structures denoted by the number of neurons in the input (descriptors), hidden, and output layers, the cutoff radii, as well as individual computational costs (μs/atom/CPU-core/step) of evaluating individually the inter-nuclear distances within $r_c$, the density-like structural descriptors, and NNs (plus the rest minor contributions), respectively. NN structures and cutoff radii are identical for EANN and PEANN.

| System | NN structure | $r_c$ (Å) | Individual costs: Distances/Descriptors/Others | |
|---|---|---|---|---|
| | | | PEANN | EANN |
| Cu | 10×10×10×1 | 4.1 | 1.2/1.9/1.8 | 1.2/3.5/1.8 |
| Ge | 15×15×15×1 | 5.0 | 1.2/3.8/2.6 | 1.3/7.8/2.5 |
| Mo | 16×16×16×1 | 5.0 | 1.7/4.2/2.8 | 1.6/7.9/2.5 |
| $H_2O$ | 33×20×20×1 | 6.3 | 3.3/23.9/6.5 | 3.3/66.3/8.0 |

Table S2: Hyperparameters of the piecewise descriptors of the PEANN Cu potential.

| Numbering | $L_{max}$ | $r_{in}$ (Å) | $r_{out}$ (Å) | α |
|---|---|---|---|---|
| 1 | 0/1 | 0.50 | 1.70 | 12.83 |
| 2 | 0/1 | 1.10 | 2.30 | 6.04 |
| 3 | 0/1 | 1.70 | 2.90 | 2.85 |
| 4 | 0/1 | 2.30 | 3.50 | 1.35 |
| 5 | 0/1 | 2.90 | 4.10 | 0.63 |

Table S3: Hyperparameters of the piecewise descriptors of the PEANN Ge potential.

| Numbering | $L_{max}$ | $r_{in}$ (Å) | $r_{out}$ (Å) | α |
|---|---|---|---|---|
| 1 | 0/1/2 | 1.00 | 2.09 | 9.35 |
| 2 | 0/1/2 | 1.73 | 2.82 | 3.76 |
| 3 | 0/1/2 | 2.45 | 3.54 | 1.51 |
| 4 | 0/1/2 | 3.18 | 4.27 | 0.61 |
| 5 | 0/1/2 | 3.91 | 5.00 | 0.24 |

Table S4: Hyperparameters of the piecewise descriptors of the PEANN Mo potential.

| Numbering | $L_{max}$ | $r_{in}$ (Å) | $r_{out}$ (Å) | α |
|---|---|---|---|---|
| 1 | 0/1 | 1.50 | 2.00 | 1.60 |
| 2 | 0/1 | 0.43 | 2.43 | 7.49 |
| 3 | 0/1 | 0.86 | 2.86 | 3.66 |
| 4 | 0/1 | 1.29 | 3.29 | 0.62 |
| 5 | 0/1 | 1.71 | 3.71 | 0.19 |
| 6 | 0/1 | 3.04 | 4.14 | 0.02 |
| 7 | 0/1 | 3.17 | 4.57 | 0.02 |
| 8 | 0/1 | 3.60 | 5.00 | 0.01 |

Table S5: Hyperparameters of the piecewise descriptors of the PEANN bulk water potential.

| Numbering | Central atom | $L_{max}$ | $r_{in}$ (Å) | $r_{out}$ (Å) | $\alpha$ |
|---|---|---|---|---|---|
| 1 | O | 0/1/2 | -0.10 | 1.10 | 12.97 |
| 2 | O | 0/1/2 | 0.43 | 1.53 | 4.45 |
| 3 | O | 0/1/2 | 0.06 | 2.06 | 1.92 |
| 4 | O | 0/1/2 | 0.89 | 2.59 | 2.97 |
| 5 | O | 0/1/2 | 1.72 | 3.12 | $1.28 \times 10^{-1}$ |
| 6 | O | 0/1/2 | 1.95 | 3.65 | $1.13 \times 10^{-1}$ |
| 7 | O | 0/1/2 | 3.78 | 4.18 | $2.96 \times 10^{-2}$ |
| 8 | O | 0/1/2 | 3.31 | 4.71 | $1.42 \times 10^{-2}$ |
| 9 | O | 0/1/2 | 3.84 | 5.24 | $6.83 \times 10^{-3}$ |
| 10 | O | 0/1/2 | 4.37 | 5.77 | $3.27 \times 10^{-3}$ |
| 11 | O | 0/1/2 | 4.90 | 6.30 | $1.57 \times 10^{-3}$ |
| 1 | H | 0/1/2 | -0.40 | 1.00 | 12.12 |
| 2 | H | 0/1/2 | -0.47 | 1.57 | 3.84 |
| 3 | H | 0/1/2 | 0.06 | 2.06 | 1.84 |
| 4 | H | 0/1/2 | 0.59 | 2.59 | 1.16 |
| 5 | H | 0/1/2 | 1.72 | 3.12 | $1.29 \times 10^{-1}$ |
| 6 | H | 0/1/2 | 2.25 | 3.65 | $6.19 \times 10^{-2}$ |
| 7 | H | 0/1/2 | 2.78 | 4.18 | $2.97 \times 10^{-2}$ |
| 8 | H | 0/1/2 | 3.31 | 4.71 | $1.42 \times 10^{-2}$ |
| 9 | H | 0/1/2 | 3.84 | 5.24 | $6.83 \times 10^{-3}$ |
| 10 | H | 0/1/2 | 4.37 | 5.77 | $3.27 \times 10^{-3}$ |
| 11 | H | 0/1/2 | 4.90 | 6.30 | $1.57 \times 10^{-3}$ |

Table S6: Hyperparameters of the descriptors of the EANN Cu potential.

| Numbering | $L_{max}$ | $r_s$ (Å) | $\alpha$ (Å$^{-2}$) |
|---|---|---|---|
| 1 | 0/1 | 0.00 | 0.22 |
| 2 | 0/1 | 0.95 | 0.22 |
| 3 | 0/1 | 1.90 | 0.22 |
| 4 | 0/1 | 2.85 | 0.22 |
| 5 | 0/1 | 3.80 | 0.22 |

Table S7: Hyperparameters of the descriptors of the EANN Ge potential.

| Numbering | $L_{max}$ | $r_s$(Å) | $\alpha$ (Å$^{-2}$) |
|---|---|---|---|
| 1 | 0/1/2 | 0.00 | 0.15 |
| 2 | 0/1/2 | 1.15 | 0.15 |
| 3 | 0/1/2 | 2.30 | 0.15 |
| 4 | 0/1/2 | 3.45 | 0.15 |
| 5 | 0/1/2 | 4.60 | 0.15 |

Table S8: Hyperparameters of the descriptors of the EANN Mo potential.

| Numbering | $L_{max}$ | $r_s$(Å) | $\alpha$ (Å$^{-2}$) |
|---|---|---|---|
| 1 | 0/1 | 0.00 | 0.43 |
| 2 | 0/1 | 0.68 | 0.43 |
| 3 | 0/1 | 1.36 | 0.43 |
| 4 | 0/1 | 2.04 | 0.43 |
| 5 | 0/1 | 2.72 | 0.43 |
| 6 | 0/1 | 3.40 | 0.43 |
| 7 | 0/1 | 4.08 | 0.43 |
| 8 | 0/1 | 4.76 | 0.43 |

Table S9: Hyperparameters of the descriptors of the EANN bulk water potential.

| Numbering | Central atom | $L_{max}$ | $r_s$(Å) | $\alpha$ (Å$^{-2}$) |
|---|---|---|---|---|
| 1 | O | 0/1/2 | 0.00 | 0.54 |
| 2 | O | 0/1/2 | 0.61 | 0.54 |
| 3 | O | 0/1/2 | 1.22 | 0.54 |
| 4 | O | 0/1/2 | 1.83 | 0.54 |
| 5 | O | 0/1/2 | 2.44 | 0.54 |
| 6 | O | 0/1/2 | 3.05 | 0.54 |
| 7 | O | 0/1/2 | 3.66 | 0.54 |
| 8 | O | 0/1/2 | 4.27 | 0.54 |
| 9 | O | 0/1/2 | 4.88 | 0.54 |
| 10 | O | 0/1/2 | 5.49 | 0.54 |
| 11 | O | 0/1/2 | 6.10 | 0.54 |
| 1 | H | 0/1/2 | 0.00 | 0.54 |
| 2 | H | 0/1/2 | 0.61 | 0.54 |
| 3 | H | 0/1/2 | 1.22 | 0.54 |
| 4 | H | 0/1/2 | 1.83 | 0.54 |
| 5 | H | 0/1/2 | 2.44 | 0.54 |
| 6 | H | 0/1/2 | 3.05 | 0.54 |
| 7 | H | 0/1/2 | 3.66 | 0.54 |
| 8 | H | 0/1/2 | 4.27 | 0.54 |
| 9 | H | 0/1/2 | 4.88 | 0.54 |
| 10 | H | 0/1/2 | 5.49 | 0.54 |
| 11 | H | 0/1/2 | 6.10 | 0.54 |

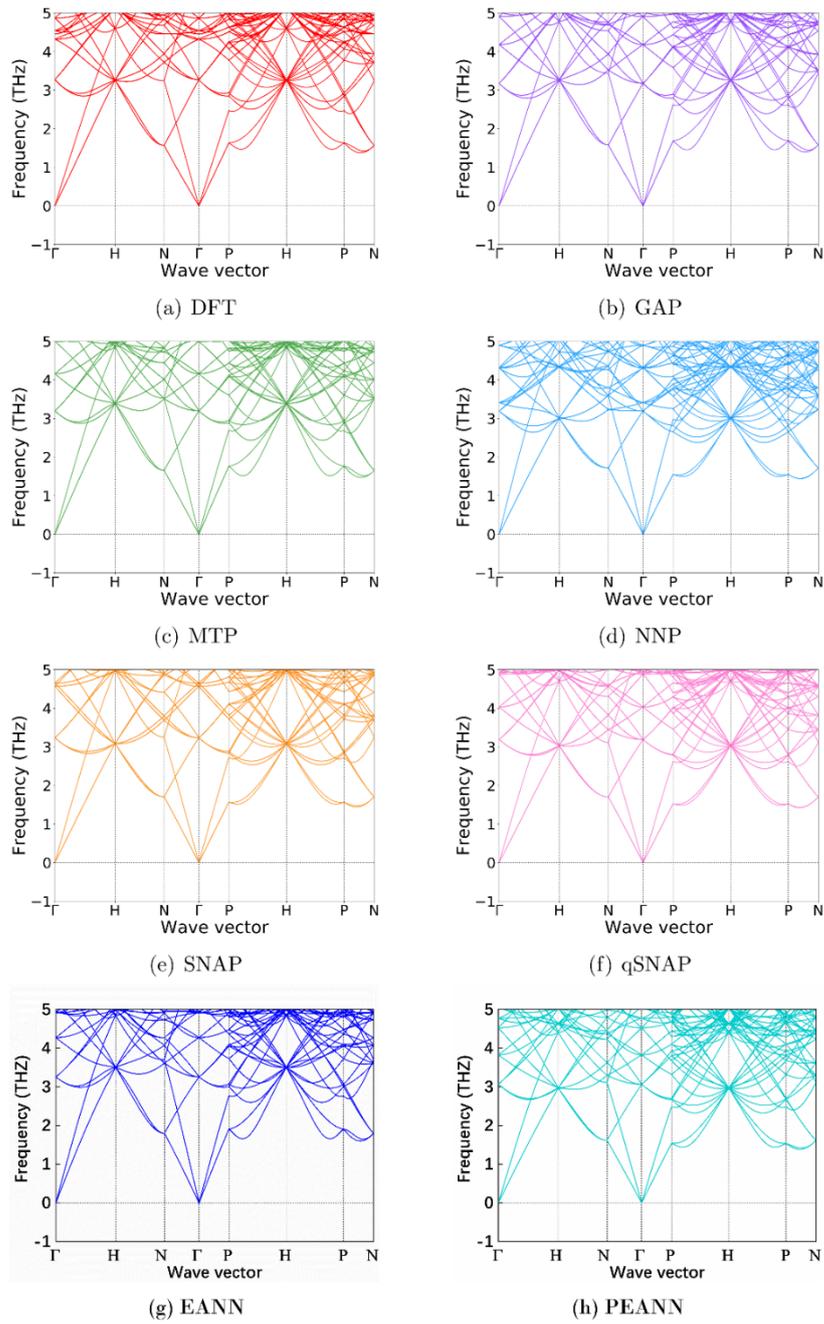

Fig. S1 Phonon spectra for 3×3×3 bulk Mo obtained by (a) DFT, (b) GAP, (c) MTP, (d) NNP, (e) SNAP, (f) QSNAP (g) EANN and (h) PEANN. Note: (a)~(f) were extracted from Ref. 12. The phonon spectra were plotted along the path through the Brillouin zone given by the high-symmetry points Γ-H-N-Γ-P-H-P-N.

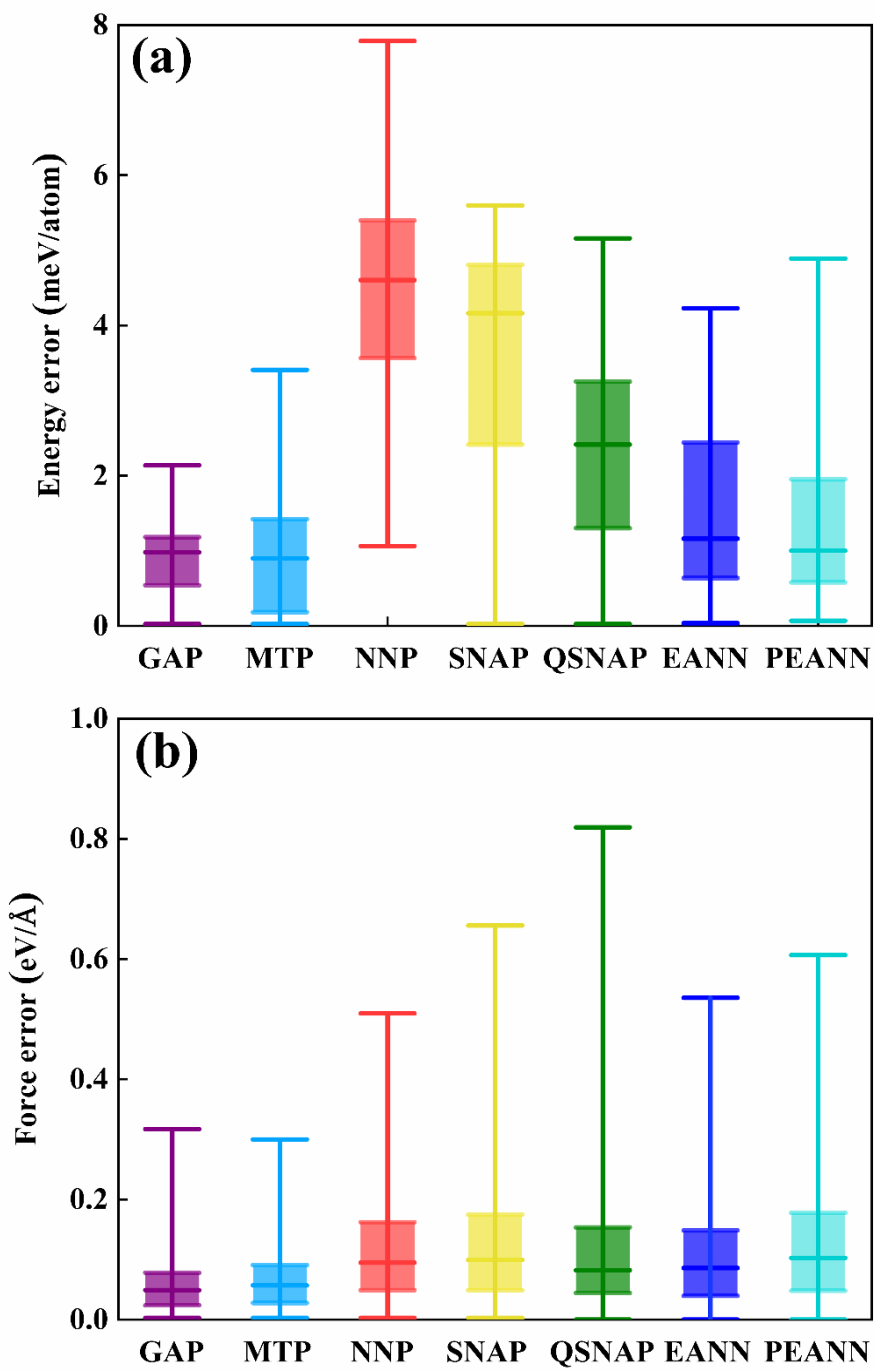

Fig. S2 Predicted error distributions of (a) energies and (b) atomic forces with 40 structures sampled from MD trajectories based on each ML model. The color filled areas are the interquartile range and the lines between them represent the median.

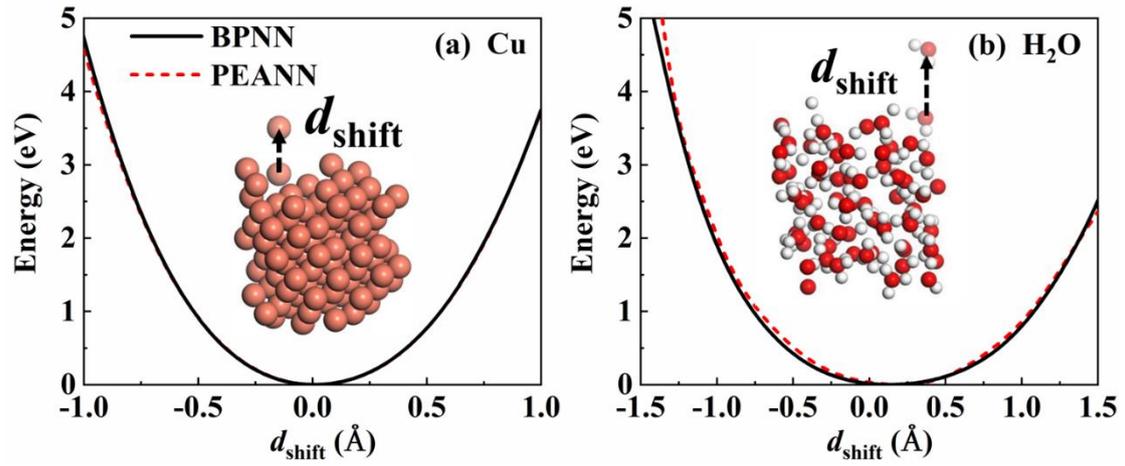

Fig. S3 Comparison of one-dimensional potential energy curves of BPNN and PEANN potentials as a function of the displacement of (a) a Cu atom and (b) a $H_2O$ molecule relative to its original position in a random configuration of Cu and water bulk structures, respectively.